\documentclass[12pt,twoside,french,english]{IEEEtran}
\usepackage[T1]{fontenc}
\usepackage{color}
\usepackage{textcomp}
\usepackage{amstext}
\usepackage{amssymb}
\usepackage{graphicx}
\usepackage{setspace}
\makeatletter
\newenvironment{lyxlist}[1]
{\begin{list}{}
{\settowidth{\labelwidth}{#1}
 \setlength{\leftmargin}{\labelwidth}
 \addtolength{\leftmargin}{\labelsep}
 }}
{\end{list}}

\onecolumn

\makeatother

\usepackage{babel}
\addto\extrasfrench{%
   \providecommand{\fg}{\ifdim\lastskip>\z@\unskip\fi~\frqq}%
}

\begin{document}

\title{A Statistical Analysis of Multipath Interference for Impulse Radio
UWB Systems}

\author{Mihai I. Stanciu$^{5}$, Stéphane Azou$^{1,3,*},$ Emanuel R\u{a}doi
$^{1,2}$ and Alexandru \c{S}erb\u{a}nescu$^{4}$, \emph{Senior Member,
IEEE\vspace{5mm}
}}

\maketitle
\begin{center}
$^{1}${\small Université Européenne de Bretagne, France.}\\
$^{2}${\small Université de Brest ; CNRS, UMR 6285 LabSTICC ; Brest,
France}\\
$^{3}${\small Ecole Nationale d'Ingénieurs de Brest; CNRS, UMR 6285
LabSTICC ; Brest, France}\\
$^{4}${\small Military Technical Academy, Bucharest, Romania}
\par\end{center}{\small \par}

\begin{center}
$^{5}${\small Freescale Semiconductor, Bucharest, Romania}
\par\end{center}{\small \par}

\begin{center}
$^{*}${\small Corresponding author (E-mail : }\texttt{\small azou@enib.fr}{\small )}
\par\end{center}{\small \par}
\begin{abstract}
In this paper, we develop a statistical characterization of the multipath
interference in an Impulse Radio (IR)-UWB system, considering the
standardized IEEE 802.15.4a channel model. In such systems, the chip
length has to be carefully tuned as all the propagation paths located
beyond this limit can cause interframe/intersymbol interferences (IFI/ISI).
Our approach aims at computing the probability density function (PDF)
of the power of all multipath components with delays larger than the
chip time, so as to prevent such interferences. Exact analytical expressions
are derived first for the probability that the chip length falls into
a particular cluster of the multipath propagation model and for the
statistics of the number of paths spread over several contiguous clusters.
A power delay profile (PDP) approximation is then used to evaluate
the total interference power as the problem appears to be mathematically
intractable. Using the proposed closed-form expressions, and assuming
minimal prior information on the channel state, a rapid update of
the chip time value is enabled so as to control the signal to interference
plus noise ratio.\end{abstract}

\section{Introduction}

Since its approval by the U.S. Federal Communications Commission (FCC)
in 2002, UWB has given rise to considerable interest in wireless communications
research community, due to its many attractive properties \cite{[Arslan06]}.
Today, IR is considered as a promising technique from an industrial
perspective \cite{[Willig08]}; in particular, it is a main candidate
solution for applications such as wireless sensor networks \cite{[Zhang09]}
due to its ability to provide joint data transmission and precise
positioning \cite{[Gezici05]}. However, a number of IR-UWB system
design challenges remain to be solved to ensure a broader use of that
technology in practice \cite{[Lampe10]}. The very fine time resolution
in IR transmission coupled with the rich multipath diversity of the
UWB channels need a careful signal and architecture design to achieve
good performances at reasonable complexity. The large path delay spread
may require sophisticated signal processing algorithms at the receiver
to cope with IFI/ISI. This can be avoided if we increase the minimum
delay between two consecutive time hopping code values, that is to
say if we decrease the chip rate (and thus the bit rate), or if we
don't make use of consecutive code elements (corresponding to a time
lag of one chip). The multiuser interference (MUI) has also to be
taken into account in system design and some schemes limiting its
impact on performances may be required. Considering the chips as adjacent
channels available for multiple users, the main cause of interference
between channels is the time spread of the channel, if it exceeds
the chip duration. In particular, it is shown in \cite{[Hu04]} that
performance of correlator-based TH-PPM and TH-BPSK receivers is severely
degraded by the MUI. Thus, given a particular channel profile, any
information on the power of interfering multipath components (MPCs)
can provide valuable assistance either for evaluating the signal to
interference plus noise ratio (SINR) and the performance in terms
of bit error rate, or for adapting the system parameters to the propagation
conditions. This problem has been investigated only in a limited number
of papers. Approximate analytical expressions for Signal to ISI power
ratio was first derived in \cite{[Piazzo05]} for various UWB transmission
formats over two paths channels. In \cite{[Deleuze05]}, a closed-form
expression of the ISI/IFI energy at the output of a rake receiver
is stated by considering the IEEE 802.15.3a statistical channel model
\cite{[Molisch09a]} with one cluster. The influence of various parameters
(time-hopping code, rake number of fingers, guard-time size) on the
performance is conducted, hence enabling a pertinent design of the
system. Even if IFI is an important issue, especially for high data
rate UWB systems, \cite{[Ahmed06]} considers that its influence can
eventually be neglected in BPSK or PPM time-hopping systems under
certain conditions : a large number of frames per symbol is required
and the considered scenario neglects ISI and multiple access (transmission
of a single symbol for a single user). Gezici \emph{et al.} established
in \cite{[Gezici07]} that there is a tradeoff between the pulse combining
gain $N_{f}$ and the pulse spreading gain $N_{c}=T_{f}/T_{c}$ to
get a minimal bit errror probability in presence of timing jitter,
for frequency-selective environments, as IFI is mitigated for larger
values of $N_{c}$ , the effect of timing jitter being mitigated by
increasing $N_{f}$. A different approach is proposed in \cite{[Witrisal08]}
to analyse the influence of channel and system parameters : by computing
the first and second order moments of the received UWB pulses, the
authors can express the intra-burst interference resulting from multipath
propagation. In \cite{[Rahman09]}, the authors derive a few analytical
expressions for IFI/ISI in pulsed direct sequence (DS) and hybrid
DS/TH UWB communications, considering a frequency selective Nakagami
fading channel. With standard Gaussian approximation of ISI/IFI, an
expression of the error probability is given for a MRC rake combiner
and the effects of various system parameters on the performances are
studied.

In this paper, we seek to express the probability density function
(PDF) of the power of the whole set of MPCs having delays exceeding
the chip length, given a particular channel profile stemming from
a prior identification step \cite{[Venkatesh07]}. We consider in
particular the statistical channel model adopted by the IEEE 802.15.4a
Task Group%
\footnote{The interest to consider this model is that it is widely studied by
the research community, due to its ability to reflect propagation
phenomena with acceptable statistical precision. However, there is
still ongoing researches to improve this model such as \cite{[Haneda12]}
where the frequency dependence of the multipath components is investigated.
Extensions of the present work could then be considered for future
research, depending on the evolution of standards.%
} \cite{[Molisch06]} to derive an analytical form of the PDF with
some approximations. Proposed method may facilitate the characterization
of a radio link in terms of SINR. Our theoretical analysis involves
the following intermediate steps to achieve the interference power
characterization : first, we derive the probability that the chip
length falls into a particular cluster of the multipath propagation
model; then, the statistics of the number of paths in a cluster is
expressed, which yields the statistics of the number of paths spread
over several contiguous clusters; finally, we obtain an approximate
relationship of the PDF of the power of all interfering MPCs using
a simplified power delay profile expression. Due to dense multipath
propagation and the non-stationary nature typical to UWB channels,
the level of interference can change very quickly and \textcolor{black}{an
adaptation of some key parameters of the modulation may be required
so as to control IFI/ISI (or multiuser interference) according to
the environment in which the system is operating. Such possibility
has been recently investigated in \cite{[Molisch09b]}, where the
frame duration is periodically updated depending on the channel state
information and SINR, which are measured using training sequences.
The development proposed here can serve as a tool for a rapid performance
assessment of a TH UWB radio link; as only a LOS/NLOS detection is
required as prior information, our result could possibly be applied
in a non-data aided scenario. }

The paper is organized as follows. Section II briefly recalls some
properties of the IEEE 802.15.4a propagation channel and specifies
the problem of statistical analysis of the interference power. Then,
in section III, we derive the statistics of the first cluster matching
the chip length and the number of MPCs is investigated in section
IV. A power delay approximation then follows in section V and a procedure
for interference power estimation is developed in section VI. A few
numerical results are then discussed before some conclusions to confirm
the pertinence of our theoretical analysis.

\section{System Model and Problem Statement}

We consider here a Time-Hopping Binary Pulse Amplitude Modulation
(BPAM) format, with the following typical expression of the transmitted
signal :

\begin{equation}
s_{tx}(t)=\sqrt{\frac{E_{s}}{N_{f}}}\sum_{p=-\infty}^{\infty}d_{\left\lfloor p/N_{f}\right\rfloor }w_{tx}\left(t-pT_{f}-c_{p}T_{c}\right),\label{eq:UWB transmitted signal model}
\end{equation}
where $E_{s}$ denotes the symbol energy and $d_{\left\lfloor p/N_{f}\right\rfloor }\in\left\{ -1,\,1\right\} $
are the binary transmitted symbols, $T_{f}$ stands for the average
period of the pulse train (also known as the frame time), the $c_{p}\in\{0,1,...,N_{c}-1\}$
represent pseudo-random code elements required both for code division
multiple access and spectral shaping purposes, the quantification
of temporal hops being controlled by the chip length $T_{c}$. The
system is designed so that the unit energy pulse $w_{tx}(t)$ is confined
within duration $T_{c}$ and we have $T_{f}=N_{c}T_{c}$. The indoor
propagation channel usually involves multiple reflections due to the
objects in the vicinity of receiver and transmitter; a classical way
to characterize such propagation conditions is to make use of the
Saleh-Valenzuela (SV) channel model \cite{[Molisch09a]} whose basic
assumption is that MPCs arrive in clusters, with complex variations
of the received power due to path loss, large scale fading and small
scale fading. The corresponding discrete-time impulse response is
usually expressed as 
\begin{equation}
h(t)=\sum_{l=0}^{L}\sum_{k=0}^{K}a_{k,l}exp(j\phi_{k,l})\delta(t\text{\textminus}T_{l}\text{\textminus}\tau_{k,l})\label{eq:ChannelImpulseResponse}
\end{equation}

where $a_{k,l}$ denotes the tap weight of the $k$-th component in
the $l$-th cluster, $\tau_{k,l}$ is the delay of the $k$-th MPC
relative to the $l$-th cluster arrival time $T_{l}$ and the phases
$\phi_{k,l}$ being uniformly distributed in the range $[0,2\pi]$;
$\delta(t)$ is the Dirac delta function.

\begin{figure}[tbh]
\begin{centering}
\includegraphics[width=11cm]{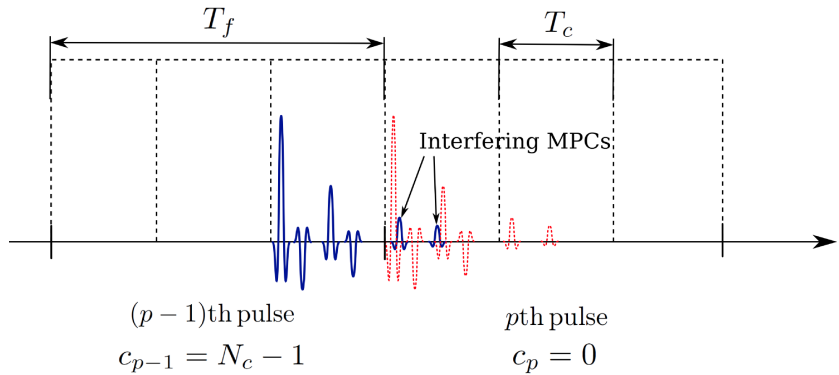}
\par\end{centering}

\caption{\label{fig:Inter-pulse-interference}Interframe interference from
the $(p-1)$th frame to the $p$th frame; in this example $N_{c}=3$
and the two consecutive pulses have a relative delay equal to $T_{c}$.}
\end{figure}

In case the maximum delay spread of the channel is larger than the
chip time, the transmitted signal corresponding to one pulse may overlap
with signals in some next frames, as shown in Fig. \ref{fig:Inter-pulse-interference},
thus causing IFI/ISI. Our objective is to derive a closed form expression
of the power lying in time bins with delays beyond the chip length,
with the aim of controlling the SINR. As will be seen in the sequel,
the problem rapidly becomes intractable due to the complexity of the
mathematical relations involved; a few approximations will then be
proposed, in such a way that the SINR is overestimated.

Many variations on the SV model have been proposed in the literature
according to the considered environment, frequency band and transmission
range. The popular IEEE 802.15.4a statistical channel model will be
considered in the following, as it is close to a realistic channel;
moreover, it is valid for UWB systems irrespective of their data rate
and their modulation format. Based on measurements and simulations
in various environements, the 802.15.4a model includes several improvements
on previously proposed statistical models : a frequency dependent
path gain is used, the number of clusters $L$ is assumed to be Poisson
distributed, ray arrival times are modeled via a mixture of Poisson
processes and different shapes of power delay profiles (PDP) are assumed
to better reflect the line-of-sight (LOS) or non-line-of-sight (NLOS)
configurations.

\section{Statistics of the first cluster matching the chip length}

In this section, we turn our attention to the probability $P\left(T_{c}\in\mathcal{C}_{\ell}\right)$
that the chip length ``falls into the $\ell$-th cluster''%
\footnote{In the sequel, the index $\ell$ will refer to the cluster containing
the chip length value whereas index $l$ will be used for any other
cluster.%
} $\mathcal{C}_{\ell}=[T_{\ell-1},T_{\ell})$, that is the probability
that $T_{\ell-1}\leq T_{c}\leq T_{\ell}$. To achieve this goal, it
is required to compute the PDF $f_{\ell}\left(x\right)=\int_{0}^{x}f_{\ell}\left(x|T_{0}\right)f\left(T_{0}\right)\mbox{d}T_{0}$
of the cluster arrival time $T_{\ell}$, supposing that $f\left(T_{0}\right)=\Lambda_{0}\mbox{exp}\left(-\Lambda_{0}T_{0}\right)$,
$T_{0}\geq0$. Using the expression $T_{\ell}=\left(\sum_{j=1}^{\ell}\Delta T_{j}+T_{0}\right)$
and considering that distributions of the cluster arrival times are
given by a Poisson processes with cluster arrival rate $\Lambda$,
i.e. \foreignlanguage{french}{$f_{\Delta T_{\ell}}\left(x\right)=\Lambda\mbox{exp}\left(-\Lambda x\right)$
where $\Delta T_{\ell}=T_{\ell}-T_{\ell-1}$, we get}

\selectlanguage{french}%
\begin{equation}
f_{\ell}\left(x\right)=\frac{\Lambda_{0}\Lambda^{\ell}\mbox{exp}\left(-\Lambda x\right)}{\left(\ell-1\right)!}\int_{0}^{x}\left(x-t\right)^{\ell-1}\mbox{exp}\left(-\left(\Lambda_{0}-\Lambda\right)t\right)\mbox{d}t.\label{eq:Forme_interm_pdf_T_k}
\end{equation}

Then, by observing that integral of the form $I_{l,b}\left(x\right)=\int_{0}^{x}\left(x-t\right)^{l}\mbox{exp}\left(-bt\right)\mbox{d}t$
satisfies the recursive relation%
\footnote{This relation results from the integration by parts $\int_{0}^{x}u'v=[uv]_{0}^{x}-\int_{0}^{x}uv'$,
with $u'=exp(-bt)$ and $v=(x-t)^{l}$.%
} $I_{l,b}\left(x\right)=\frac{x^{l}}{b}-\frac{l}{b}I_{l-1,b}\left(x\right)$
with $b=\Lambda_{0}-\Lambda$, we can easily show that
\begin{eqnarray}
f_{\ell}\left(x\right) & = & \frac{\Lambda_{0}\Lambda^{\ell}\mbox{exp}\left(-\Lambda x\right)}{\left(\ell-1\right)!}\frac{\left(-1\right)^{\ell-1}(\ell-1)!}{(\Lambda_{0}-\Lambda)^{\ell}}\label{eq:f_l(x)}\\
 &  & \times\left(\sum_{i=0}^{\ell-1}\frac{\left(-bx\right)^{\ell-1-i}}{\left(\ell-1-i\right)!}-\mbox{exp}\left(-(\Lambda_{0}-\Lambda)x\right)\right)\nonumber 
\end{eqnarray}

Before computing the probability of interest \foreignlanguage{english}{$P\left(T_{c}\in\mathcal{C}_{\ell}\right)$,
let us define }$z=T_{c}$, $x=T_{\ell-1}$ and $y=\Delta T_{\ell}$
to simplify the mathematical notations. Then, we can rewrite \foreignlanguage{english}{$P\left(T_{c}\in\mathcal{C}_{\ell}\right)$
as 
\begin{eqnarray}
P\left(z\in\mathcal{C}_{\ell}\right) & = & \mbox{P}\left(z-y<x\leq z\right)\label{eq:P(z in C_l)}\\
 & = & \int_{0}^{z}\left(\int_{z-y}^{z}f_{\ell}\left(x\right)\mbox{d}x\right)\Lambda\mbox{exp}\left(-\Lambda y\right)\mbox{d}y\nonumber \\
 &  & +\int_{z}^{\infty}\left(\int_{0}^{z}f_{\ell}\left(x\right)\mbox{d}x\right)\Lambda\mbox{exp}\left(-\Lambda y\right)\mbox{d}y\nonumber 
\end{eqnarray}
}

\selectlanguage{english}%
To proceed further, it is required to express integral of the type
\foreignlanguage{french}{$J_{m,\Lambda}\left(T_{c}\right)=\int_{T_{c}}^{\infty}x^{m}\mbox{exp}\left(-\Lambda x\right)\mbox{d}x$
and $\overline{J}_{m,\Lambda}\left(T_{c}\right)=\int_{0}^{T_{c}}x^{m}\mbox{exp}\left(-\Lambda x\right)\mbox{d}x$,
where $\Lambda\in\mathbb{R}_{+}^{*}$ and $m\in\mathbb{N}$. It can
be easily verified that }$J_{m,\Lambda}\left(T_{c}\right)+\overline{J}_{m,\Lambda}\left(T_{c}\right)=\frac{m!}{\Lambda^{m+1}}$
and $J_{m,\Lambda}\left(T_{c}\right)=\frac{\mbox{exp}\left(-\Lambda T_{c}\right)}{\Lambda^{m+1}}\sum_{p=0}^{m}\frac{m!\left(T_{c}\Lambda\right)^{p}}{p!}$.

Finally, from the relations above we obtain the following result after
a few mathematical calculations :
\begin{equation}
\mbox{P}\left(T_{c}\in\mathcal{C}_{\ell}\right)=\frac{\Lambda_{0}\Lambda^{\ell}}{\left(\Lambda-\Lambda_{0}\right)^{\ell+1}}\mbox{e}^{-\Lambda T_{c}}R_{\ell}\left(-\left(\Lambda_{0}-\Lambda\right)T_{c}\right)\label{eq:pdf_first_cluster}
\end{equation}

where \foreignlanguage{french}{$R_{\ell}\left(u\right)$ stands for
the rest of the $\ell$-th order Taylor series expansion of the function
$\mbox{e}^{u}$, that is
\begin{equation}
R_{\ell}\left(u\right)=e^{u}-\sum_{i=0}^{\ell}\frac{u^{i}}{i!}\label{eq:taylor_series_rest}
\end{equation}
}

\selectlanguage{french}%
Note that we assumed throughout the previous development that $T_{c}$
is larger than the first cluster delay $T_{0}$. In case we have $T_{c}<T_{0}$,
the interference corresponds to all the received pulse power. This
event has the probability $\mbox{P}\left(\left\{ T_{c}<T_{0}\right\} \right)=\mbox{e}^{-\Lambda_{0}T_{c}}$.

\selectlanguage{english}%

\section{On the number of interfering MPCs}

This section is devoted to the estimation of the number of MPCs that
can cause interference, that is the overall number of components located
beyond the chip length. So, for any cluster index $l\in\mathbb{N}$
, we need to compute first the probability $P(n_{i})$ for each subsequent
cluster $\mathcal{C}_{i}$, $l+1\leq i\leq n,$ where $n_{i}$ denotes
the number of MPCs belonging to the $i$-th cluster. The distribution
associated to the whole set of interfering clusters will then be easily
obtained in a second step. 

In order to simplify our development, we will consider the approximation
that there is no inter-cluster interference; hence the delay $\tau_{m,i}$
of the $m$-th MPC relative to the $i$-th cluster arrival time $T_{i-1}$
belongs to the interval \foreignlanguage{french}{$\left[T_{i-1},\, T_{i}\right)$}.
A second simplification is assumed for the ray arrival times; from
\cite{[Molisch06]} we know that they can be modelled with mixtures
of two Poisson processes with mixture probability $\beta$ and arrival
rates $\{\lambda_{1},\lambda_{2}\}$ being determined experimentally
for various environments. As in the classical SV model, we will adopt
a Poisson process for the ray arrival times, so that the distribution
of the delay difference $\tau_{l+1,i}-\tau_{l,i}$ of any two adjacent
MPCs in cluster $\mathcal{C}_{i}$ takes the form $f(\tau_{l+1,i}-\tau_{l,i})=\lambda exp(-\lambda(\tau_{l+1,i}-\tau_{l,i}))$,
where the value of the parameter $\lambda$ is computed by minimizing
the mean squared error between simplified and original models, for
a given radio environment (known values for $\beta,\,\lambda_{1}$,
and $\lambda_{2}$).

For any cluster $\mathcal{C}_{i}$, it can be shown (see Appendix)
that the probability $P(n_{i})$ that the number of MPCs in the cluster
is $n_{i}$, is\foreignlanguage{french}{
\begin{equation}
\mbox{P}\left(n_{i}\right)=\frac{\lambda^{n_{i}}\Lambda}{\left(\lambda+\Lambda\right)^{n_{i}+1}},\,\,\, n_{i}\geq0\label{eq:P(ni)}
\end{equation}
}

\selectlanguage{french}%
Then, we can characterize the number of MPCs within clusters \{$\mathcal{C}_{i};\, i=\ell,\ell+1,...,L$\},
where $L$ is the total number of clusters and supposing that $T_{c}\in\mathcal{C}_{\ell}$.
This can be achieved by summing $r+1$ indepedent and identically
discrete random variables, each with probability (\ref{eq:P(ni)}),
where $r=L-\ell$. The resulting probability $\mbox{P}_{r}\left(n\right)$
can be computed by recursion%
\footnote{Note that this probability depends on the number $(r+1)$ of clusters
beyond $\mathcal{C}_{\ell}$, with $T_{c}\in\mathcal{C}_{\ell}$.%
}, starting with $\mbox{P}_{1}\left(n\right)=\sum_{k=0}^{n}\mbox{P}_{0}\left(k\right)\mbox{P}_{0}\left(n-k\right)$
and $\mbox{P}_{0}\left(n\right)$ being expressed as in (\ref{eq:P(ni)})
:

\begin{equation}
\mbox{P}_{r}\left(n\right)=\frac{\lambda^{n}}{\left(\lambda+\Lambda\right)^{n}}\left(\frac{\Lambda}{\left(\lambda+\Lambda\right)}\right)^{r+1}\frac{\left(n+r\right)!}{n!r!}.\label{eq:All_clusters_nr_of_MPCs}
\end{equation}

\selectlanguage{english}%

\section{Power Delay Profile Approximation}

The Power Delay Profile (PDP) is defined as the squared magnitude
of the channel impulse response, averaged over the small-scale fading.
In the frame of IEEE 802.15.4a it is expressed as
\begin{equation}
E\left\{ \left|a_{k,l}\right|^{2}\right\} \propto\Omega_{l}exp\left(-\tau_{k,l}/\gamma_{l}\right)\label{eq:pdp}
\end{equation}

where the integrated energy $\Omega_{l}$ over the $l$th cluster
follows an exponential decay and the intra-cluster decay time constant
$\gamma_{l}$ depends linearly on the arrival time of the cluster.
This model makes the closed-form derivation of the interfering power
estimate intractable; so we will adopt the following first approximation
for the PDP : 
\begin{equation}
\Omega(t)=\Omega_{c}exp\left(-\frac{t}{\Gamma}\right)\label{eq:PDP_original}
\end{equation}

where $t$, $\Omega_{c}$ and $\Gamma$ denote the path delay, the
integrated energy of the cluster and the intra-cluster decay time
constant, respectively.

Concerning the small-scale fading, it can be shown that the $i$-th
path relative to the $l$-th cluster has its power distributed as\foreignlanguage{french}{
\begin{equation}
f\left(x|\,\Omega_{il}\right)=\frac{1}{\Gamma\left(m\right)}\left(\frac{m}{\Omega_{il}}\right)^{m}x^{m-1}\mbox{exp}\left(-\frac{mx}{\Omega_{il}}\right),\label{eq:MPC-power-pdf}
\end{equation}
}

\selectlanguage{french}%
where $\Omega_{il}$ stands for the PDP of the considered path, $m$
is the Nakagami $m$-factor of the small-scale amplitude distribution
and $\Gamma\left(m\right)$ is the gamma function.

For CM1/CM2 channel models, the above relation can be simplified by
considering a mean value%
\footnote{As the random variable $m$ follows a lognormal distribution, the
mean value is expressed as $\mu_{m}=exp(m_{0}+\hat{m}_{0}^{2}/2),$where
the parameters ($m_{0},\hat{m_{0})}$ have specified values depending
on the considered environment \cite{[Molisch06]}.%
} $m\cong2$ for the parameter $m$; in this case $\Gamma(m)=1$ and
we get
\begin{equation}
f\left(x|\,\Omega_{il}\right)=\left(\frac{2}{\Omega_{il}}\right)^{2}x.\mbox{exp}\left(-\frac{2x}{\Omega_{il}}\right),\label{eq:MPC-power-pdf-simplified}
\end{equation}

which represents a $\Gamma(\alpha,\theta)$ distribution with $\alpha=2$
and $\theta=\frac{\Omega_{il}}{2}$.

\selectlanguage{english}%
Now, we need to express the mean value of the PDP corresponding to
the whole set of MPCs located beyond the chip length :
\begin{equation}
\Omega_{0}=E\left[\Omega_{T_{c}}\left(\tau_{i}\right)\right]\label{eq:Omega_0}
\end{equation}

\selectlanguage{french}%
where $\Omega_{T_{c}}\left(\tau_{i}\right)=\Omega\left(\tau_{i}\right)$,
for $\tau_{i}\geq T_{c}$ and zero elsewhere, $\tau_{i}$ denoting
the path delay.\textbf{ }We propose to derive an approximate expression
of $\Omega_{0}$ by picking $n$ uniformly spaced samples of the original
distribution (\ref{eq:PDP_original}) in the interval $\left[T_{c},\, T_{L}\right)$,
where $T_{L}$ corresponds to the upper bound of the $L$-th cluster
: 
\begin{equation}
\Omega_{0}\left(L\right)=\mbox{exp}\left(-\frac{T_{c}}{\Gamma}\right)\frac{1}{n}\sum_{i=0}^{n-1}\left[\mbox{exp}\left(\frac{T_{c}}{\Gamma}\right)\mbox{exp}\left(-\frac{T_{N}}{\Gamma}\right)\right]^{i/n},\label{eq:Omega_0(L)}
\end{equation}

which has the equivalent form
\begin{equation}
\Omega_{0}\left(L\right)=\frac{1}{n}\frac{g_{L}-g_{0}}{\left(g_{L}/g_{0}\right)^{1/n}-1},\label{eq:Omega0-NoApproximation}
\end{equation}

where $g_{0}=\mbox{exp}\left(-T_{c}/\Gamma\right)$ and $g_{L}=\mbox{exp}\left(-T_{L}/\Gamma\right)$. 

An additional simplification can be achieved for sufficiently large
value of $n$ by considering that
\begin{equation}
\left[\left(g_{L}/g_{0}\right)^{1/n}-1\right]/\left(1/n\right)\simeq\mbox{ln}\left(g_{L}/g_{0}\right)\label{eq:Approx-Omega0-Large-n}
\end{equation}

Hence, we finally obtain the following approximation for the PDP :
\begin{equation}
\Omega_{0}\left(L\right)\cong\left\{ \begin{array}{cc}
\Gamma\left[\mbox{exp}\left(-T_{c}/\Gamma\right)-\mbox{exp}\left(-T_{L}/\Gamma\right)\right]\\
/\left(T_{L}-T_{c}\right), & \,\, T_{L}>T_{c},\\
\mbox{exp}\left(-T_{c}/\Gamma\right), & \,\, T_{L}=T_{c}.
\end{array}\right..\label{eq:Approx-Omega0}
\end{equation}

\begin{figure}[tbh]
\selectlanguage{english}%
\begin{centering}
\includegraphics[scale=0.7]{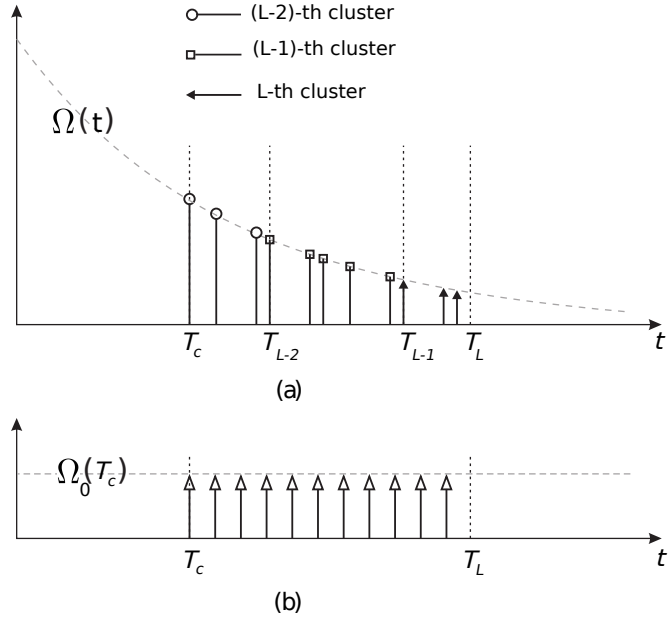}
\par\end{centering}

\selectlanguage{french}%
\caption{\selectlanguage{english}%
\label{fig:PDP-Approximation-Principle}Proposed approximation of
the PDP\selectlanguage{french}%
}
\end{figure}

As can be seen in Fig. \ref{fig:PDP-Approximation-Principle}, considering
(\ref{eq:Approx-Omega0-Large-n}) generally gives a good approximation
of (\ref{eq:Omega0-NoApproximation}), for various values of $L$
and $T_{c}$, with an overestimation of the true value. Concerning
the number of clusters, it can be assumed a Poisson distribution \cite{[Molisch06]}
\begin{equation}
P_{L}\left(L\right)=\frac{\overline{L}^{L}e^{-\overline{L}}}{L!},\label{eq:Prob_nombre_de_clusters}
\end{equation}
where $\overline{L}$ denotes the average number of clusters. 

\selectlanguage{english}%
We know also that the last cluster delay has its PDF given by $f\left(T_{L}\right)=\frac{T_{L}^{L-1}\Lambda^{L}\mbox{exp}\left(-\Lambda T_{L}\right)}{\left(L-1\right)!},$
hence\foreignlanguage{french}{
\begin{equation}
f_{T_{L}}\left(T_{L}|T_{L}\geq T_{c}\right)=\left\{ \begin{array}{cc}
T_{L}^{L-1}\mbox{exp}\left(-\Lambda T_{L}\right)\\
/\left(\int_{T_{c}}^{\infty}t^{L-1}\mbox{exp}\left(-\Lambda t\right)\mbox{dt}\right), & T_{L}\geq T_{c},\\
0, & T_{L}<T_{c}.
\end{array}\right.,\label{eq:pdf-TN}
\end{equation}
}

From this last equation, we can then derive a closed form expression
of the mean PDP, that will be used in the next section, devoted to
interference power estimation :
\begin{eqnarray}
\overline{\Omega}_{0}\left(L\right) & =E\left[\Omega_{0}(L)\right]\,\,\,\,\,\,\,\,\,\,\,\,\,\,\,\,\,\,\,\,\,\,\,\,\,\,\,\,\,\,\,\,\,\,\,\,\,\,\,\,\,\,\,\,\,\,\,\,\,\,\,\,\,\,\,\,\,\,\,\,\,\,\,\,\,\,\,\nonumber \\
\cong & \left(\Gamma\int_{T_{c}}^{\infty}\frac{\mbox{exp}\left(-T_{c}/\Gamma\right)-\mbox{exp}(-x/\Gamma)}{x-T_{c}}x^{L-1}\mbox{exp(-\ensuremath{\Lambda}x)}\mbox{dx}\right)\label{eq:mean-Omega0}\\
 & /\left(\int_{T_{c}}^{\infty}t^{L-1}\mbox{exp}\left(-\Lambda t\right)\mbox{dt}\right)\nonumber 
\end{eqnarray}

\section{Interference power estimation}

To begin with, let us recapitulate below the intermediate results
derived until now :
\begin{lyxlist}{00.00.0000}
\item [{-}] first, we derived in section III the probability that the chip
length ``falls into a particular cluster'' of the multipath propagation
model;
\item [{-}] then, the statistics of the number of paths in any cluster
was expressed in section IV, together with the statistics of the total
number of paths located beyond the chip duration;
\item [{-}] a simplified power delay profile expression has then be derived
in section V. 
\end{lyxlist}
We are now going to derive an approximate relationship of the PDF
of the power of all interfering MPCs. Our approach relies on the summation
of a number $n$ of r.v. each following the PDF (\ref{eq:MPC-power-pdf-simplified}),
with the factor $\theta=\Omega_{0}(L)$ computed as (\ref{eq:mean-Omega0}).
Hence, for a given number of clusters $N$, the resulting PDF is expressed
as, for $x>0$,\foreignlanguage{french}{
\begin{equation}
g\left(x|\,\Omega_{0}\left(L\right)\right)=\sum_{n=1}g\left(x|\,\Omega_{0}\left(L\right),n\right)\cdot\mbox{P}\left(n|L\right)\label{eq:interf_pdf_function_of_N}
\end{equation}
}

\selectlanguage{french}%
where 

\begin{equation}
g\left(x|\,\Omega_{0}\left(L\right),n\right)=\left(\frac{2}{\Omega_{0}\left(L\right)}\right)^{2n}\frac{x^{2n-1}}{\left(2n-1\right)!}\cdot\mbox{exp}\left(-\frac{2x}{\Omega_{0}\left(L\right)}\right)\label{eq:interf_pdf_function_of_n_N}
\end{equation}

and the number of paths with \foreignlanguage{english}{delays} exceeding
$T_{c}$ being determined as (under the assumption that each cluster
contains at least one path)
\begin{equation}
\mbox{P}\left(n|L\right)=\sum_{k=0}^{L-1}\mbox{P}\left(n|k,L\right)\cdot\mbox{P}\left(k\right),\label{eq:Prob_nombre_trajets_N_n}
\end{equation}

The above expression can be computed by taking advantage of previous
developments; for example, in case of a LOS environment (and considering
that $T_{0}=0$), we get
\begin{equation}
\begin{array}{c}
\mbox{P}\left(n|k,L\right)=\left(\frac{\lambda}{\lambda+\Lambda}\right)^{n-L+k}\left(\frac{\Lambda}{\left(\lambda+\Lambda\right)}\right)^{L-k}\\
\,\,\,\,\,\,\,\,\,\,\,\,\,\,\,\,\,\,\,\,\,\,\,\,\times\frac{\left(n-1\right)!}{\left(n-L+k\right)!\left(L-k-1\right)!},\,\,\,\,\,\,\,\,\,\,\,\,\,\,\forall n\geq1
\end{array}\label{eq:Prob_nombre_trajets_N_n_k}
\end{equation}

and
\begin{equation}
\mbox{P}\left(k\right)=\frac{\left(\Lambda T_{c}\right)^{k}\mbox{exp}\left(-\Lambda T_{c}\right)}{k!}.\label{eq:Prob_Ck_LOS}
\end{equation}
To complete our analysis, we need also to consider the case $x=0$,
which results in a discrete part of the PDF :
\begin{eqnarray}
g\left(x=0|\,\Omega_{0}\left(L\right)\right) & = & \mbox{P}\left(n=0|L\right)\label{eq:g(x=00003D0|Omega_0(L))}\\
 & = & \sum_{k=L}^{\infty}\mbox{P}\left(k\right)\nonumber 
\end{eqnarray}

And finally, the interference power is obtained as
\begin{equation}
g\left(x\right)=\sum_{L=1}^{\infty}g\left(x|\,\Omega_{0}\left(L\right)\right)P_{L}\left(L\right).\label{eq:Interference_globale}
\end{equation}

No compact analytical expression of this PDF can be obtained due to
the high complexity of the terms involved; however, we can observe
that the proposed result can easily be implemented to get an estimate
at very limited computational cost. Note also that our development
remains valid for both LOS and NLOS environments; in the latter case,
the \foreignlanguage{english}{probability that the chip length falls
into the $k$th cluster has the alternative expression }
\begin{equation}
\mbox{P}\left(k\right)=\frac{\left(\Lambda T_{c}\right)^{k+1}\mbox{exp}\left(-\Lambda T_{c}\right)}{(k+1)!}.\label{eq:P_k_NLOS}
\end{equation}

\begin{figure}[h]
\selectlanguage{english}%
\begin{centering}
\includegraphics[width=15cm]{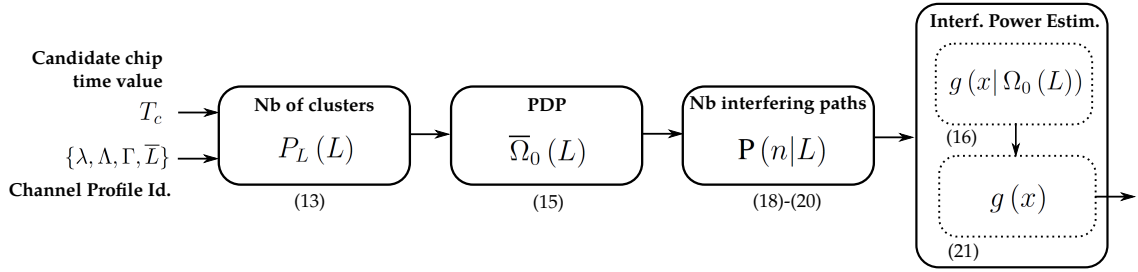}
\par\end{centering}

\selectlanguage{french}%
\caption{\label{fig:algorithm-block-diagram}Block diagram of the proposed
algorithm for interference power estimation}
\end{figure}

\selectlanguage{english}%
The proposed algorithm for interference power estimation is summarized
in the form of a block diagram in Fig. (\ref{fig:algorithm-block-diagram}).

\selectlanguage{french}%

\section{Simulation results}

\begin{figure}[h]
\selectlanguage{english}%
\begin{centering}
\includegraphics[scale=0.7]{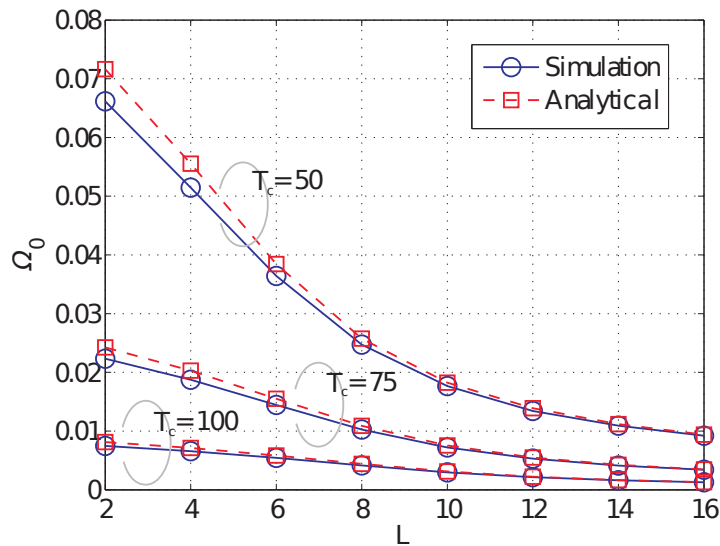}
\par\end{centering}

\selectlanguage{french}%
\caption{\label{fig:Mean-value-of-Omega0}Mean value of $\Omega_{0}\left(L\right)$
for various chip length values (CM1 channel model;$10^{4}$ MC simulations)}
\end{figure}

\selectlanguage{english}%
As mentioned before, our theoretical analysis involves a few approximations
to get the statistics of the interference power, due to the high complexity
of some mathematical relations :
\begin{enumerate}
\item A Poisson process is adopted for modelling the ray inter-arrival times,
instead of a mixture of two Poisson processes as recommended in the
frame of the IEEE 802.15.4a channel model;
\item A simplified exponential expression (\ref{eq:PDP_original}) is considered
for the mean power of the different paths (PDP); we do not take into
account the possible different values of the integrated energy and
decay time constant for distinct clusters;
\item The computation of the mean PDP associated to MPCs located beyond
the chip duration is achieved owing to a simple sampling scheme.
\end{enumerate}
A great number of Monte-Carlo simulations have been conducted using
the Matlab implementation \cite{[Molisch04]} of the IEEE 802.15.4a
channel model to verify the pertinence of our approach. A few illustrations
are given hereafter to show the impact of various approximations on
the statistics. The case of CM1 channel model is considered here,
but it should be noticed that the same procedure could be applied
in another radio environement. Firstly, we can examine the error resulting
from the proposed uniform sampling for the computation of the mean
PDP (\ref{eq:Approx-Omega0-Large-n}). As can be seen in Fig. \ref{fig:Mean-value-of-Omega0},
the computed value of $\overline{\Omega}_{0}\left(L\right)$ tends
to match the true value when the number of clusters $L$ increases
and the error decreases for larger chip length. Also, it can be clearly
observed that our approximation yields an overestimation of the mean
PDP, which is particularly important from the point of view of applications.

\selectlanguage{french}%

\begin{figure}[h]
\selectlanguage{english}%
\begin{centering}
\includegraphics[width=12cm]{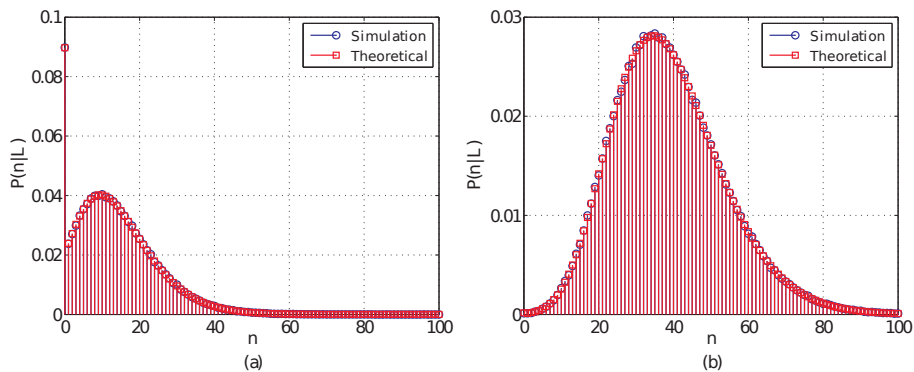}
\par\end{centering}

\selectlanguage{french}%
\caption{\selectlanguage{english}%
\label{fig:pdf-of-the-number-of-paths}PDF of the number of paths
with delays exceeding the chip length $T_{c}=50$ (CM1 channel model;
$10^{5}$ MC runs) : (a) $L=5$ clusters (b) $L=10$ clusters\selectlanguage{french}%
}
\end{figure}

\selectlanguage{english}%
We can now consider the PDF (\ref{eq:Prob_nombre_trajets_N_n}) of
the number of paths with delays larger than the chip length, which
plays a central role in estimating the interference power. As shown
in Fig. \ref{fig:pdf-of-the-number-of-paths}, there is almost a perfect
match between the values obtained through simulations and the values
obtained by numerical evaluation of analytical expressions.  Evidently,
it can be seen that the numbers of MPCs increases with the number
of clusters.

\selectlanguage{french}%

\begin{figure}[h]
\selectlanguage{english}%
\begin{centering}
\includegraphics[width=14cm]{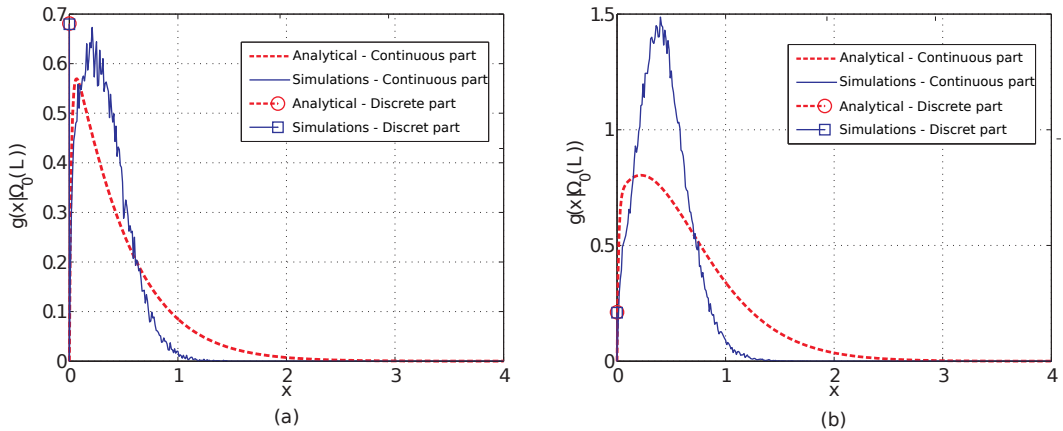}
\par\end{centering}

\selectlanguage{french}%
\caption{\selectlanguage{english}%
\label{fig:g_x_Omega0_L_Tc50}PDF \foreignlanguage{french}{\textcolor{black}{$g\left(x|\,\Omega_{0}\left(L\right)\right)$for
CM1 channel model, $T_{c}=50$ : (a) $L=2$ clusters (b) $L=4$ clusters}}\selectlanguage{french}%
}
\end{figure}

\begin{figure}[h]
\selectlanguage{english}%
\begin{centering}
\includegraphics[width=14cm]{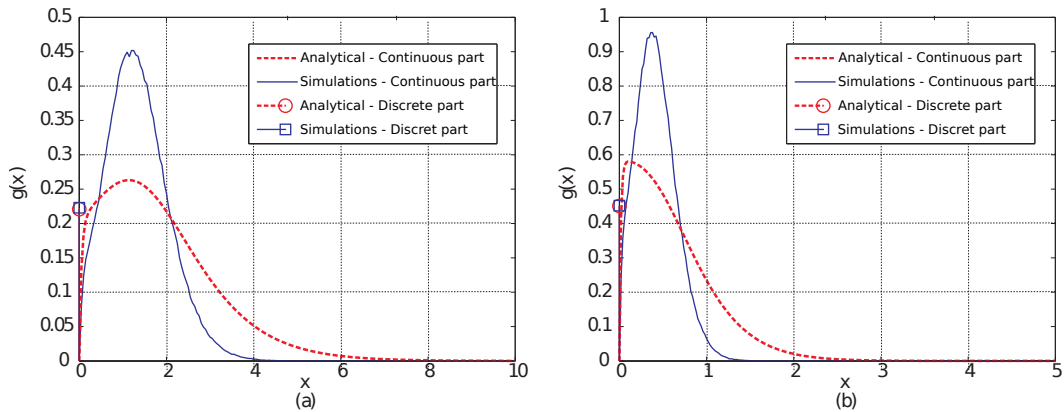}
\par\end{centering}

\selectlanguage{french}%
\caption{\selectlanguage{english}%
\label{fig:g_x}PDF of the interference power $g(x)$\foreignlanguage{french}{
\textcolor{black}{for CM1 channel model ($10^{5}$MC runs) : (a) $T_{c}=25$
(b) $T_{c}=50$}}\selectlanguage{french}%
}
\end{figure}

Figure \foreignlanguage{english}{(\ref{fig:g_x_Omega0_L_Tc50}) illustrates
the PDF }\textcolor{black}{$g\left(x|\,\Omega_{0}\left(L\right)\right)$}\foreignlanguage{english}{
resulting from summing $n$ r.v., each one following (\ref{eq:MPC-power-pdf-simplified}),
once the mean PDP associated to interfering MPCs has been computed.
Then, Fig. \ref{fig:g_x} shows the PDF of the interference power
for two distinct chip time values. The difference between the ``true''
distribution (estimated through MC simulations) and the PDF derived
from our method comes from the approximations 2) and 3) explained
above. However, this limited statistical precision is not an obstacle
for practical applications since a low error is noticed for the two
first moments of the distribution. Throughout numerous simulations
we always noticed that the ``true'' mean value is systematically
upper bounded by the value obtained via our method. A poorer fit is
observed regarding the second order moment, but again the true variance
is upper bounded by the computed value for $g(x)$.}

\begin{figure}[h]
\selectlanguage{english}%
\begin{centering}
\includegraphics[width=7cm]{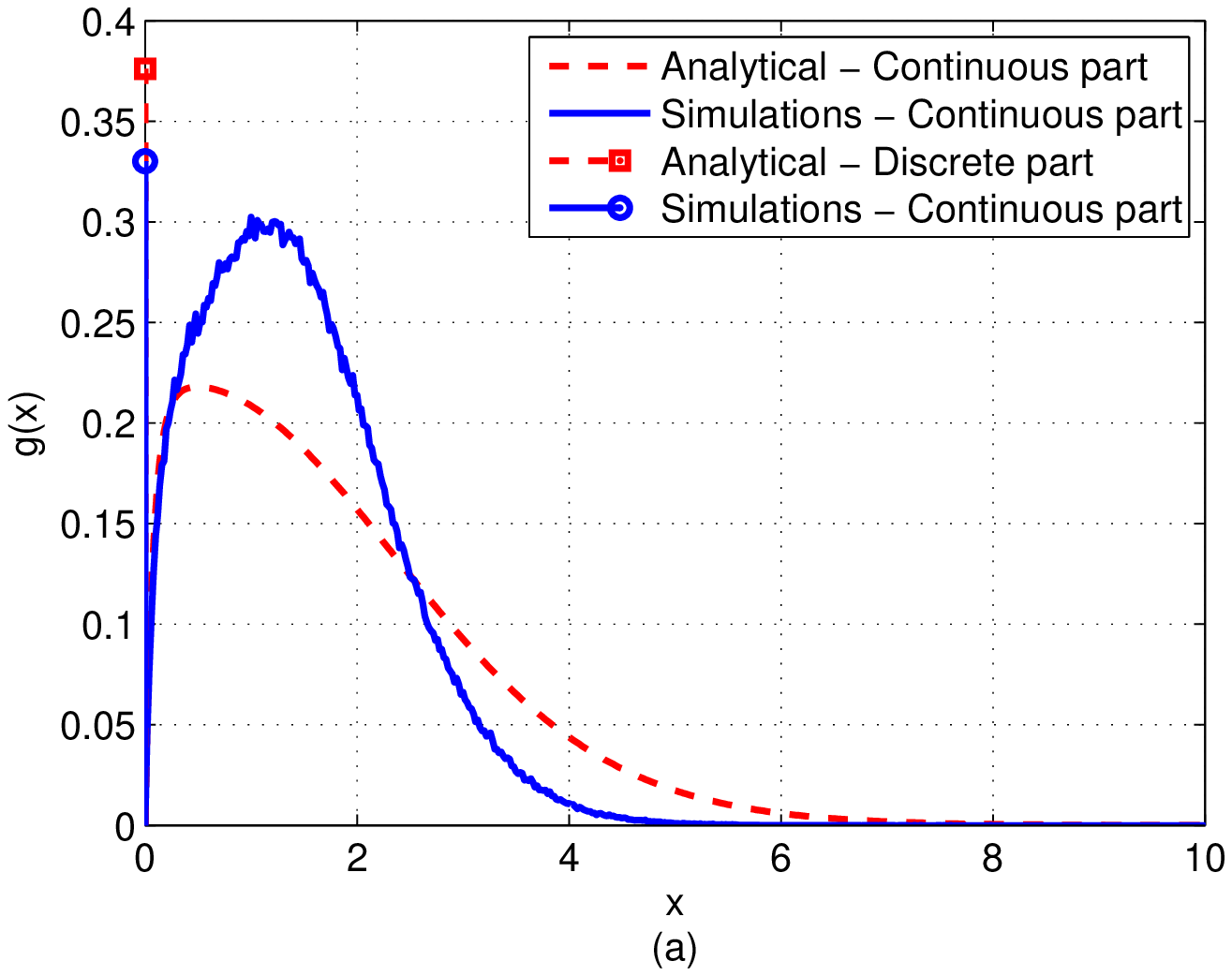}\includegraphics[width=9cm]{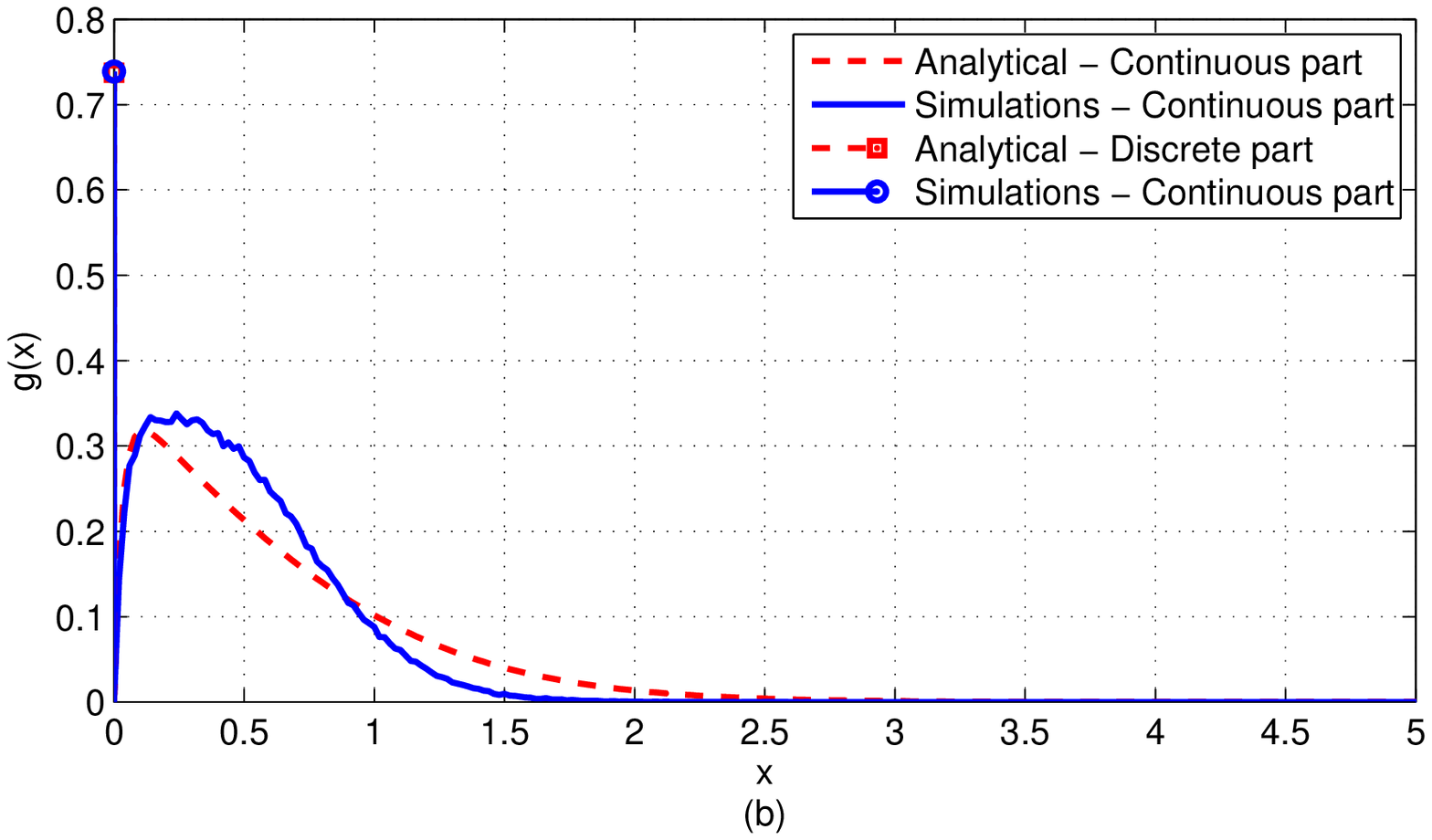}
\par\end{centering}

\selectlanguage{french}%
\caption{\selectlanguage{english}%
\label{fig:g_x-CM2}PDF of the interference power $g(x)$\foreignlanguage{french}{
\textcolor{black}{for CM2 channel model ($10^{5}$MC runs) : (a) $T_{c}=25$
(b) $T_{c}=50$}}\selectlanguage{french}%
}
\end{figure}

\begin{figure}[h]
\selectlanguage{english}%
\begin{centering}
\includegraphics[width=7cm]{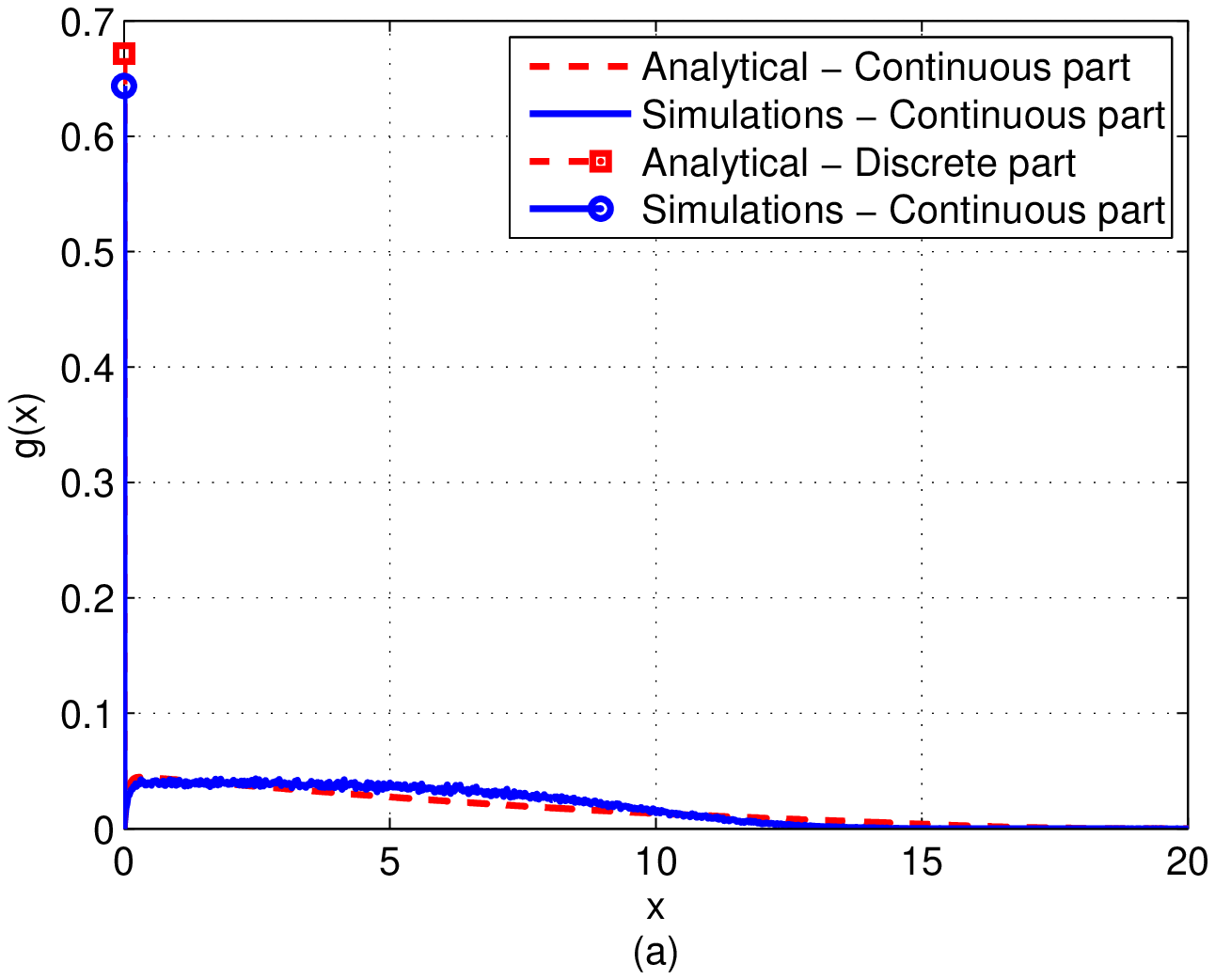}\includegraphics[width=7cm]{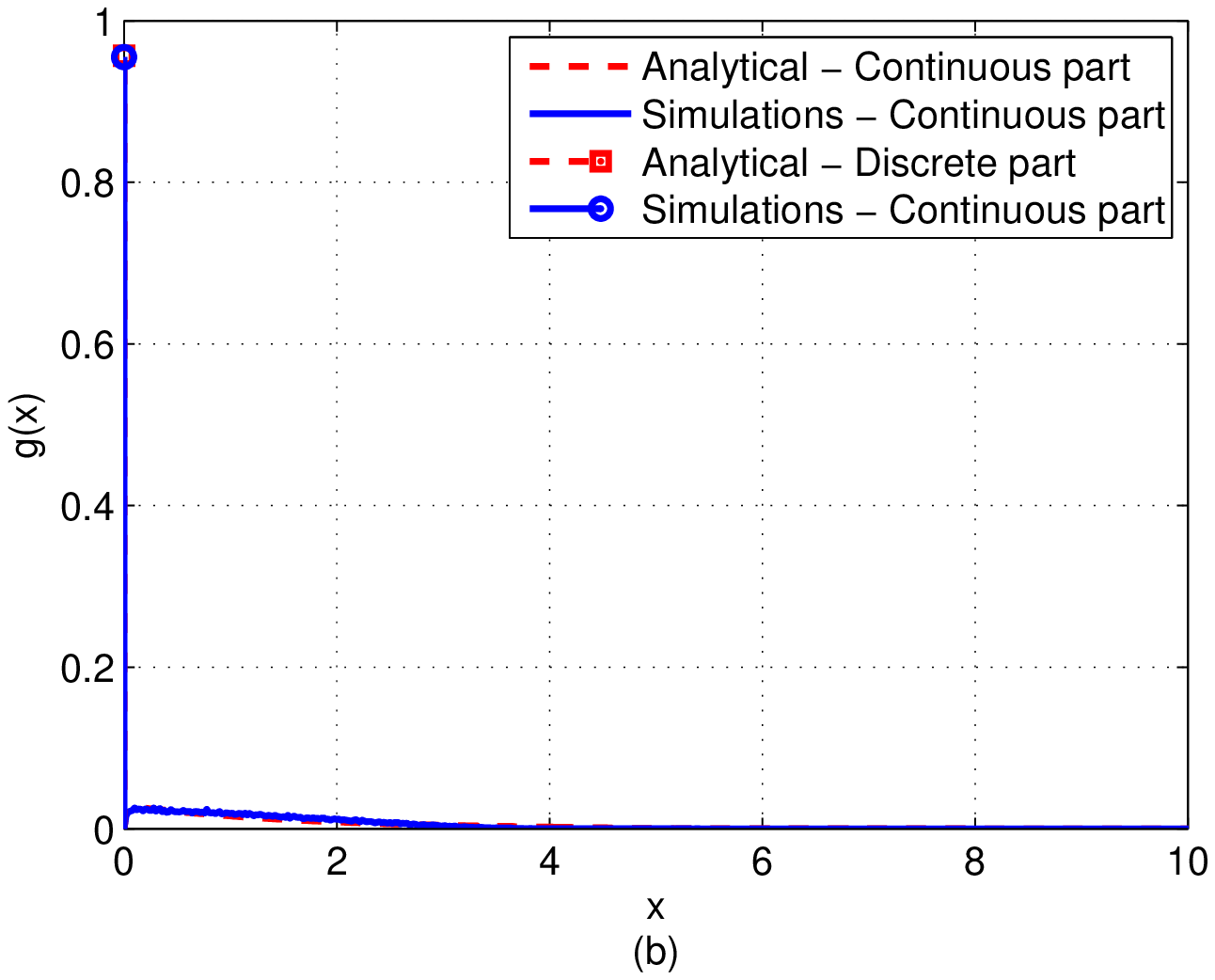}
\par\end{centering}

\selectlanguage{french}%
\caption{\selectlanguage{english}%
\label{fig:g_x-CM4}PDF of the interference power $g(x)$\foreignlanguage{french}{
\textcolor{black}{for CM4 channel model ($10^{5}$MC runs) : (a) $T_{c}=25$
(b) $T_{c}=50$}}\selectlanguage{french}%
}
\end{figure}

Additional simulations have been conducted in NLOS radio environments
to evaluate the pertinence of the proposed algorithm. For CM2/CM4
channels (NLOS residential/office), we obtained the PDFs depicted
in Fig. \ref{fig:g_x-CM2} \& \ref{fig:g_x-CM4}, for the same candidate
values of the chip time. The results appear to be acceptable, with
a good precision regarding the first two moments.

\selectlanguage{english}%

\section{Conclusions}

A novel approach for the statistical characterization of the multipath
interference in IR-UWB systems has been developed in this paper. In
the frame of the IEEE 802.15.4a channel model, we derived a theoretical
analysis which aims to compute the power corresponding to all multipath
components located beyond the chip length. Since the proposed approach
requires minimal knowledge on the channel state, it can be helpful
for real time adaptation of modulation parameters so as to limit the
signal to interference plus noise ratio. Our method relies on the
following key developments : first, we derived the probability that
the chip length falls into a particular cluster of the multipath propagation
model; then, the statistics of the number of paths spread over several
contiguous clusters have been computed in closed-form; finally, we
obtained an approximate relationship of the PDF of the power of all
interfering MPCs using a simplified power delay profile expression.
Numerous Monte-Carlo simulations have been carried out to verify the
pertinence of our results; although these simulations revealed a significant
gap between the true interference power distribution and that obtained
through our approach, the first two computed moments are very close
and upper-bound the true ones. This is a very useful result because
they represent an information that helps us to control the SINR level
and to ensure an effective functioning of the IR-UWB system in real-world
scenarios. An experimental validation of these theoretical results
is also planned for the near future using an UWB platform recently
acquired by our research team. Although some measured data acquired
with this platform is already available, it could not be used in the
framework of this paper because the conditions required for accurately
matching the IEEE 802.15.4a environment have not yet been met. An
already scheduled upgrade of our experimental UWB platform will lead
to increased performance, especially in terms of bandwidth, and will
enable an appropriate and meaningful comparison of the theoretical
results presented in this paper to those that will be obtained from
measured data.

\appendix{\textbf{\emph{Appendix A}}\emph{ - Proof of (\ref{eq:P(ni)})}}

For any cluster $\mathcal{C}_{i}$, let us define the following notations:
$x=\tau_{n_{i},i}-\tau_{0,i}$, $y=\tau_{(n_{i}+1),i}-\tau_{n_{i},i}$
and $z=\Delta T_{i}$. The probability that the number of MPCs in
the cluster is $n_{i}$ can then be written as\foreignlanguage{french}{
\begin{equation}
\mbox{P}\left(n_{i}\right)=\int_{0}^{\infty}P\left(n_{i}|z\right)f\left(z\right)\mbox{d}z\label{eq:P(n_i)}
\end{equation}
where $f(z)=\Lambda\mbox{exp}\left(-\Lambda z\right)$ is the PDF
of the cluster arrival times and with the conditional probability
\begin{eqnarray}
\mbox{P}\left(n_{i}|z\right) & = & \mbox{P}\left(z-y<x\leq z\right)\label{eq:P(n_i|z)}\\
 & = & \int_{0}^{z}\left(\int_{z-y}^{z}f(x)\mbox{d}x\right)\lambda\mbox{exp}\left(-\lambda y\right)\mbox{d}y\nonumber \\
 &  & +\int_{z}^{\infty}\left(\int_{0}^{z}f(x)\mbox{d}x\right)\lambda\mbox{exp}\left(-\lambda y\right)\mbox{d}y\nonumber 
\end{eqnarray}
}

\selectlanguage{french}%
Considering the simplified model of the ray arrival times, it can
be easily seen that 
\begin{equation}
f(x)=\frac{x^{n_{i}-1}}{\left(n_{i}-1\right)!}\lambda^{n_{i}}\mbox{exp}\left(-\lambda x\right)\label{eq:f(x)}
\end{equation}

Therefore we get, after a few algebra steps, 
\begin{equation}
\mbox{P}\left(n_{i}|z\right)=\frac{\left(\lambda z\right)^{n_{i}}}{n_{i}!}\mbox{exp}\left(-\lambda z\right)\label{eq:P(n_i|z) 2}
\end{equation}

which finally yields (\foreignlanguage{english}{\emph{\ref{eq:P(ni)}}}).
\selectlanguage{english}%

\end{document}